\documentclass[twocolumn,aps]{revtex4}

\usepackage{amsmath}
\usepackage{epsfig}
\usepackage{graphicx}
\usepackage{color}

\begin{document}

\title{Dynamic heterogeneities above and below the mode-coupling temperature. Evidence of a dynamic crossover}
\author{Elijah Flenner and Grzegorz Szamel}
\affiliation{Department of Chemistry, Colorado State University, Fort Collins, CO 80523}
\date{\today}                                           
\begin{abstract}
We examine dynamic heterogeneities in a model glass-forming fluid, a binary harmonic sphere mixture, 
above and below the mode-coupling temperature $T_c$. We calculate the
ensemble independent susceptibility $\chi_4(\tau_\alpha)$ and the dynamic correlation length $\xi_4(\tau_\alpha)$
at the $\alpha$-relaxation time $\tau_\alpha$. We also examine in detail the temperature dependence of 
$\tau_\alpha$ and the diffusion coefficient $D$. For higher temperatures we find that 
the standard Stokes-Einstein relationship, $D \sim \tau_\alpha^{-1}$, holds, but at lower temperatures a
fractional Stokes-Einstein relationship, $D \sim \tau_\alpha^{-\sigma}$  with $\sigma = 0.65$, emerges.
By examining the relationships between $\tau_\alpha$, $D$, $\chi_4(\tau_\alpha)$, and $\xi_4(\tau_\alpha)$
we determine that the emergence of the fractional Stokes-Einstein relationship is accompanied by 
a dynamic crossover from
$\tau_\alpha \sim e^{k_2 \xi_4}$ at higher temperatures
to $\tau_\alpha \sim e^{k_1 \xi_4^{3/2}}$ at lower temperatures. 
\end{abstract}

\maketitle
\section{Introduction}

The fundamental understanding of the drastic slowing down of a super-cooled liquid's dynamics and the eventual formation of a glass
remains an open question after decades of research. One of the challenges is finding and
understanding the universal features of glassy systems related to the dramatic increase of the 
relaxation time $\tau_\alpha$ (or viscosity $\eta$) as the liquid is cooled. 
This challenge is complicated by the observation that several dynamic regimes exist.
The dynamic regimes are usually identified by examining 
the temperature dependence of the relaxation time $\tau_\alpha$ or
the diffusion coefficient $D$ as a function of temperature $T$. The validity of different
theories, and the range at which these theories are applicable, are often tested through the fitting of $\tau_\alpha$ 
to the predictions of the theories.  

The mode-coupling theory of G\"otze and collaborators \cite{Gotze1991} in its original form predicted that upon supercooling 
the relaxation time diverges as a power law at the so-called mode-coupling temperature $T_c$, 
$\tau_\alpha \sim (T-T_c)^{-\gamma}$.  While it was quickly recognized that a true power-law divergence is absent, 
for many liquids there is a temperature range where a power law fits the temperature dependence of the relaxation time very well
over about three decades of the change of the relaxation time \cite{Berthier:2011hs}. For lower temperatures and longer relaxation times
deviations from the power-law divergence are found in both experiments \cite{Gotze:1992mp} 
and simulations \cite{Kob:1994,Kob:1995,Kob:1995pr,Szamel:2004ep}. These deviations are usually interpreted as the signature 
of a new relaxation mechanism involving activated dynamics \cite{com1}.

Several interesting observations have been made about liquids' dynamics within the temperature
range where mode-coupling like power law fits work well. This temperature range,  
which we will call hereafter the mode-coupling regime, is well accessible to simulational studies \cite{Andersen:2005ta}. 
We note that many phenomena occurring in this temperature range have been interpreted 
using the so-called potential energy landscape \cite{Heuer:2008}. Thus, this temperature region is
also referred to as the landscape influenced regime \cite{Sastry:1998,Andersen:2005ta}.
At present, the significance of the fact that the simulational results can be interpreted or rationalized
in terms of two rather different theoretical scenarios is unclear. 

One observation is the emergence of dynamic heterogeneities upon approaching the mode-coupling temperature, and
this feature has been extensively studied in simulations \cite{Berthier:2011hs,DHbook,Ediger:2000ti,Flenner2009,
Gebremichael:2001ha,Lacevic:2002bi,Lacevic:2003vq,Vogel:2004wg,Flenner2010prl,Flenner:2011ht,Karmakar:2010prl}. 
Briefly, there are subsets of particles that move either much faster or much slower than would be expected on the basis of a Gaussian
distribution of particles' displacements. Moreover, these subsets form clusters and the average size of these cluster increases upon
supercooling. Extensive work has been done to determine quantitatively the
characteristic size of dynamic heterogeneities. One way to approach this problem is to analyze the so-called four-point 
structure factor \cite{Lacevic:2003vq,Lacevic:2002bi,4point}, 
which is the structure factor calculated using the initial positions of the particles that have moved less than a 
certain distance during a specified time period (as discussed later, the distance is typically a fraction of the particle diameter and the
time is often the relaxation time). 
An analysis of this structure factor results in a dynamic correlation length $\xi_4$, which we
will refer to as the four-point dynamic correlation length. The length $\xi_4$ is found to increase with decreasing temperature. 
Thus, the increase of the dynamic correlation length
correlates with the increase in the relaxation time. It was found \cite{Karmakar:2010prl,Flenner2010prl}, 
that large systems need to be simulated for many relaxation times for an
accurate determination of $\xi_4$. The large computational effort required to perform such studies resulted in that 
the size of dynamic heterogeneities was mostly examined above the mode-coupling temperature.

Another interesting observation is the violation of the Stokes-Einstein
relation \cite{Mapes:2006ww,Ediger:2000ti,Swallen:2003tp}. 
For a liquid in its stable thermodynamic state 
it is generally found that the Stokes-Einstein relation holds and thus the diffusion coefficient is inversely proportional to 
the relaxation time, $D \sim \tau_\alpha^{-1}$. For strongly supercooled liquids this relationship breaks down 
and is often replaced by the so-called fractional Stokes-Einstein relation, $D \sim \tau_\alpha^{-\sigma}$ with $\sigma < 1$. 
The Stokes-Einstein relation violation is often rationalized in terms of the emergence of dynamic heterogeneities \cite{Ediger:2000ti}.

Recent simulations also suggest that there is some sort of dynamical crossover 
between the onset of supercooling and the mode-coupling temperature. Berthier \textit{et al.}\ \cite{Berthier:2012kd} reported
interesting changes in the dependence of the relaxation time on system size for a variety of systems, including a 
binary harmonic-sphere mixture, at temperatures around and above $T_c$.
They argued that these changes can be rationalized within the theoretical framework of the 
Random First Order Theory (which is usually assumed to include the mode-coupling theory) \cite{Biroli:2012book}. 
The latter approach allows for the competition between mode-coupling and activated
dynamics. Berthier \textit{et al.}\ argued that these dynamics are impacted by the finite system
size in different ways and this results in the change in the finite size effects in the vicinity of $T_c$.

The results of the finite size effects study were foreshadowed by an earlier dynamic correlation length study of 
the same binary harmonic-sphere mixture. 
Kob \textit{et al.}\  \cite{Kob:2011cd} froze a set of particles randomly selected out of an equilibrium configuration to form a 
stationary wall, and examined the dynamics of the particles that were still allowed to move as
a function of the distance $z$ perpendicular to the wall. By analyzing the $z$ dependence of the relaxation time, they extracted
a dynamic correlation length $\xi^{\mathrm{dyn}}$. If we need to distinguish $\xi^{\mathrm{dyn}}$ from other dynamic lengths
we will refer to it as the point-to-set dynamic correlation length. Kob \textit{et al.}\ 
found that this length initially increased, but then decreased around $T_c$. 

Initially it was believed that the decrease of $\xi^{\mathrm{dyn}}$ indicated a decrease in the
characteristic size of dynamic heterogeneities. However, by using a direct method 
to calculate a dynamic length scale $\xi_4$, which, as discussed above, is associated with 
the size of dynamic heterogeneities, we showed that dynamic heterogeneities continue to
grow below $T_c$ \cite{Flenner2012np}. We observed, however, that there is a change in the relationship between 
$\tau_\alpha$ and $\xi_4$ within the mode-coupling regime. Upon
supercooling $\tau_\alpha \sim e^{k_2 \xi_4}$, but a crossover
to $\tau_\alpha \sim e^{k_1 \xi_4^{3/2}}$ was found at a temperature slightly above $T_c$.  

In this paper we provide some details of calculations reported in Ref.  \cite{Flenner2012np} and we thoroughly examine 
the arguments for the change in the relationship between $\tau_\alpha$ and $\xi_4$. The latter crossover
remained hidden in previous simulations of a hard-sphere liquid \cite{Flenner:2011ht,Flenner2010prl} 
due to the somewhat narrower range 
of $\tau_\alpha$ and $\xi_4$ accessible in those simulations. However, 
the crossover becomes apparent after examining 
the relationship between the relaxation time and
the characteristic size and strength of dynamic heterogeneities combined with
the analysis of the Stokes-Einstein relation violation. 

The paper is organized as follows. In Sec.~\ref{sec:sim} we briefly describe the simulations
performed for this work. In Sec.~\ref{sec:dynamics} we examine single-particle
dynamics and the violation of the Stokes-Einstein relation. In particular, we characterize in some detail the 
temperature dependence of the the relaxation time and
the diffusion coefficient and compare them with the predictions of the mode-coupling theory, which allows us to  
identify the mode-coupling regime. While this identification is relatively straightforward, we found that there is some freedom and 
therefore some ambiguity regarding the identification of the mode-coupling temperature.
We also look at two other possible
functional forms for the temperature dependence of the relaxation time
and the diffusion coefficient for temperatures where the mode-coupling power 
law fits are no longer appropriate. 
In Sec.~\ref{sec:length} we study the four-point structure factor. We use it to extract the
so-called dynamic susceptibility and the dynamic correlation 
length. We examine the relationship of the latter quantities to the average dynamics, quantified by the relaxation time. We show that the emergence of 
a fractional Stokes-Einstein relation found for low temperatures is consistent with 
a crossover in the relationship between the relaxation time and
the dynamic correlation length.
We summarize the results in Sec.~\ref{sec:summary}. 
We discuss variants of the procedure used to identify the mode-coupling temperature 
and provide details of the calculation of the dynamic susceptibility in two Appendices.

\section{Simulation Details}\label{sec:sim}
We simulated a 50:50 mixture of harmonic spheres. The temperature and density dependence of this system's relaxation time
has been investigated before \cite{Berthier:2009ep,Berthier:2009}.  
This system was also used by Kob \textit{et al.} to
study the temperature dependence of the point-to-set dynamic and static correlation lengths \cite{Kob:2011cd}.
The interaction potential is given by
\begin{eqnarray}\label{eq:potential}
V_{nm}(r) & = & \frac{\epsilon}{2}\left(1-\frac{r}{\sigma_{nm}}\right)^2  \mbox{ if r} \le \sigma_{nm} \nonumber \\
&=& 0  \mbox{ if r} > \sigma_{nm}.
\end{eqnarray}  
For our binary mixture $\sigma_{22} = 1.4 \sigma_{11}$ and $\sigma_{12} = 1.2\sigma_{11}$, where the diameters are chosen
to inhibit crystallization. Our results are presented in reduced units where 
$\sigma_{11}$ is the unit of length, $\sqrt{m \sigma_{11}^2/\epsilon}$ is the
unit of time, and $10^{-4} \epsilon$ is the unit of temperature. We studied the system at a density $\rho = N/V = 0.675$. We ran simulations 
consisting of $N = N_1 + N_2 = 10,000$ particles for $30 \ge T \ge 6$, $N=40,000$ particles for $6 \ge T \ge 5$, and $N=100,000$ 
particles for $T=5$. At some temperatures we run simulations for several different $N$
to check for finite size effects. No finite size effects were observed in the simulations presented in this work.
We note that finite size effects reported by Berthier \textit{et al.}  \cite{Berthier:2012kd} were found in systems much smaller than those
investigated in our work. 

At all temperatures, we performed simulations in the $NVE$ ensemble with a time step of 0.02 using the velocity Verlet algorithm. For
$T=5$, the lowest temperature studied, we found significant energy drift after several hundred million time steps. Thus, at $T=5.5$ and $5$ 
we ran $NVT$ simulations using a Nose-Hoover thermostat with a time step of 0.06. 
The simulations were run using LAMMPS code \cite{lammps,lammpsurl}, which was modified to include the harmonic sphere
potential. 

For $T \ge 5.5$ and
every system size we ran four production runs of at least 100$\tau_\alpha$ (the $\alpha$ relaxation time, $\tau_\alpha$, is defined in
Sec.~\ref{sec:dynamics}). Note that this includes four $NVT$ simulations and four $NVE$ simulations
at $T=5.5$. For $T=5$ we ran four $NVE$ simulations for $10\tau_\alpha$ with $N=100,000$ and $40,000$ particles, and 
four $NVT$ simulations for $100\tau_\alpha$ with $N=40,000$ particles, and three $NVT$ simulations
for $100 \tau_\alpha$ with $N=100,000$. 

At every temperature we ran over $90 \tau_\alpha$ in an $NVT$ ensemble to equilibrate the system.
To ensure that the $T=5$ simulations were equilibrated, two of the 40,000 particle simulations were equilibrated for 200$\tau_\alpha$ and
these simulations produced statistically the same results as the ones equilibrated for 100$\tau_\alpha$. Note that 
for $T=5$ simulations of 100$\tau_\alpha$
with a time step of 0.06 requires $1.4\times10^9$ time steps, thus, including equilibration, the $T=5$ runs ranged from  $2.8\times10^9$ to 
$4.2\times10^9$ time steps. 
Additionally, to check for equilibration we examined two and four-point time-dependent correlation functions 
for aging. We found that the four-point susceptibility $\chi_4$ calculated directly from the
$NVE$ and $NVT$ simulations (for which we will later use symbols $\chi_4|_{NVE}$ and $\chi_4|_{NVT}$) 
is sensitive to aging as well as energy drift. We found no evidence of aging in our production runs. 
  
Finally, to check if the Nose-Hoover thermostat introduced any artifacts, we compared the results of the $NVE$ and the $NVT$ simulations
at $T=5.5$ for runs of 100$\tau_\alpha$ in both ensembles. 
No measurable differences were found. Additionally, at $T=5$ we compared an $NVE$ simulation that was ten times shorter
than the $NVT$ simulations. The results (e.g. relaxation times and four-point structure factors) 
agreed to within error. We should point out that poor choice of $NVT$ simulation parameters can lead to artifacts; see Appendix B for more details.

\section{Single Particle Dynamics and the Mode-Coupling Crossover}\label{sec:dynamics}
In this section we examine several aspects of single particle dynamics in 
super-cooled liquids. First, we examine the overlap function, and we use it to obtain the $\alpha$-relaxation time $\tau_\alpha$. 
We compare the temperature dependence of the overlap function with that of the self-intermediate scattering function. 
We also examine the mean-square displacement $\left<\delta r^2(t)\right>$, 
and calculate the diffusion coefficient $D$ from the long time behavior of the mean-square displacement. 
We examine mode-coupling-like power laws and identify the mode-coupling regime, \textit{i.e.} the temperature range where these power 
laws are a reasonable description of the data. 
In this section we discuss fits of the $\alpha$-relaxation time and the diffusion coefficient that assume the same mode-coupling
temperature for both quantities. In Appendix A we briefly discuss fits that do not use this constraint. 
Finally, we examine the Stokes-Einstein relation and find a crossover from 
the Stokes-Einstein relation to 
a fractional Stokes-Einstein relationship, which works well at low temperatures. The temperature 
at which the relationship between $D$ and $\tau_\alpha$ changes from $D \sim \tau_\alpha^{-1}$ to
$D \sim \tau_\alpha^{-\sigma}$ plays an important role in the discussion in Sec.~\ref{sec:length}.  
In this section we always analyze
quantities pertaining to the whole system (\textit{i.e.} involving both small and large particles). In Appendix A we briefly 
discuss quantities pertaining to the individual components. 

We begin by examining the the overlap function 
\begin{equation}
F_o(t) = N^{-1} \left< \sum_n w_n(t) \right>,
\end{equation}
where 
$w_n(t) = \Theta(|\mathbf{r}_n(t) - \mathbf{r}_n(0)| - a) $, $\mathbf{r}_n(t)$ is the position of particle $n$ at a time $t$, 
$\Theta$ is Heaviside step function, and, as discussed above, the sum runs over all particles of the system. 
The microscopic overlap function $\sum_n w_n(t)$ selects particles that have moved less than 
a distance $a$ over a time $t$. We fix $a=0.3$ so that the
decay time of $F_o(t)$ is approximately equal to the decay time of the self-intermediate scattering 
function $F_s(q;t)$, 
\begin{equation}\label{Fsall}
F_s(q;t) = N^{-1} \left< \sum_n e^{-i \mathbf{q} \cdot (\mathbf{r}_n(t) - \mathbf{r}_n(0))} \right>
\end{equation} 
for $q$ corresponding to the peak of the total static structure factor, $q=6.1$. In Eq. (\ref{Fsall}) the sum runs over all particles of the system. 
The overlap function and the self-intermediate scattering function encode similar information about the dynamics of our system.
For a large system, calculating the former function requires significantly less computational effort.
In addition, as discussed in the next section, we quantify dynamic
heterogeneities in terms of correlations of  the microscopic overlap function pertaining to individual particles. For these reasons, 
we define the relaxation time in terms of the overlap function. Specifically, the $\alpha$-relaxation time $\tau_\alpha$ is the time
at which the overlap function decays to $e^{-1}$, $F_o(\tau_\alpha) = e^{-1}$. 

\begin{figure}
\includegraphics[width=3.2in]{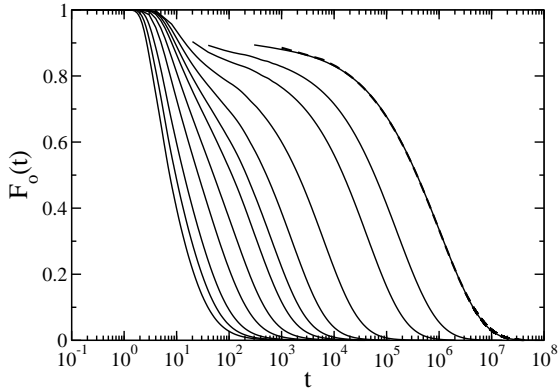}
\caption{\label{fsover}The overlap function $F_o(t)$ for $T=30$, 25, 20, 15, 12, 10, 9, 8, 7, 6 , 5.5, and 5, 
listed from left to right. The dashed line is a stretched exponential, $A_o e^{-(t/\tau_o)^{\beta_o}}$, fit to $F_o(t)$ at $T=5$
where $\beta_o = 0.53$.}
\end{figure}

Shown in Fig.~\ref{fsover} is $F_o(t)$ for $T=30$, 25, 20, 15, 12, 10, 9, 8, 7, 6, 5.5, and 5. 
The features found in $F_o(t)$ are similar to what has been seen in $F_s(q;t)$ calculated using only large particles \cite{Berthier:2009ep,Berthier:2009}
and the total $F_s(q;t)$ showed below. At high temperatures the 
overlap function agrees with what is expected from a Gaussian distribution of displacements 
(note that the high temperature limit of the overlap function can be expressed in terms of the error function and is different from
a simple exponential).  
At low temperatures a plateau develops followed by a slow, broad decay to zero. The low-temperature 
final decay is well described by a stretched exponential $A_o e^{-(t/\tau_o)^{\beta_o}}$ with 
$0.6 > \beta_o > 0.52$ for temperatures between 8 and 5.  

\begin{figure}
\includegraphics[width=3.2in]{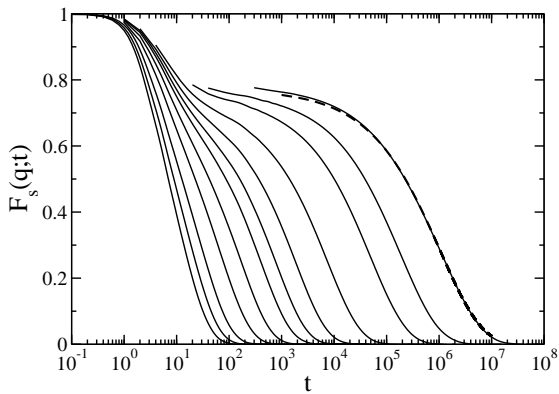}
\caption{\label{fs}The self-intermediate scattering function $F_s(q;t)$ for $q=6.1$ and $T=30$, 25, 20, 15, 12, 10, 9, 8, 7, 6, 5.5, and 5,
listed from left to right. The dashed line is a stretched exponential, $A_s e^{-(t/\tau_s)^{\beta_s}}$, fit to $F_s(q;t)$ at $T=5$
where $\beta_s = 0.55$.}
\end{figure}

The large number of particles used in our simulations makes the calculation of the intermediate correlation functions computationally 
expensive. Thus, we calculated the total self-intermediate scattering function $F_s(q;t)$ defined in Eq.~\eqref{Fsall} for a single wave-vector only. 
For $F_s(q;t)$ calculated using all particles, we used the wave-vector 
corresponding to the peak of the total structure factor, $q=6.1$. 
Shown in Fig. \ref{fs} 
is $F_s(q;t)$ for $T=30$, 25, 20, 15, 12, 10, 9, 8, 7, 6, 5.5, and 5. At high temperatures the self-intermediate scattering function
exhibits an approximately exponential decay. As was the case with the overlap function, at lower temperatures
the self-intermediate scattering function develops an intermediate plateau followed by a slow, broad decay. The low-temperature 
final decay is well described by a stretched exponential $A_s e^{-(t/\tau_s)^{\beta_s}}$ with 
$0.77 > \beta_s > 0.55$ for temperatures between 12 and 5. 

\begin{figure}
\includegraphics[width=3.2in]{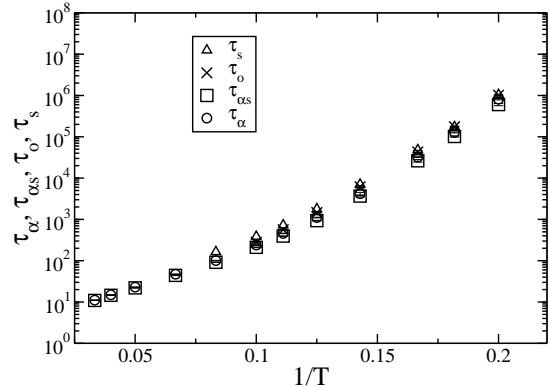}
\caption{\label{taus} Relaxation times $\tau_\alpha$, $\tau_{\alpha s}$, $\tau_o$, and $\tau_s$
plotted versus the inverse temperature.}
\end{figure}

In Fig. \ref{taus} we compare the temperature dependence of the $\alpha$-relaxation time defined above and the more usual
$\alpha$-relaxation time $\tau_{\alpha s}$ defined in terms of the decay of the self-intermediate scattering function, 
$F_s(q;\tau_{\alpha s}) = e^{-1}$. Both relaxation times exhibit very similar temperature dependence. In the same figure we also show
characteristic times obtained from the stretched exponential fits to the final decay of the overlap function, $\tau_o$ and the 
self-intermediate scattering function, $\tau_s$. These times can only be determined at low enough temperatures. Again, they exhibit a
temperature dependence similar to that of the $\alpha$-relaxation time.

\begin{figure}
\includegraphics[width=3.2in]{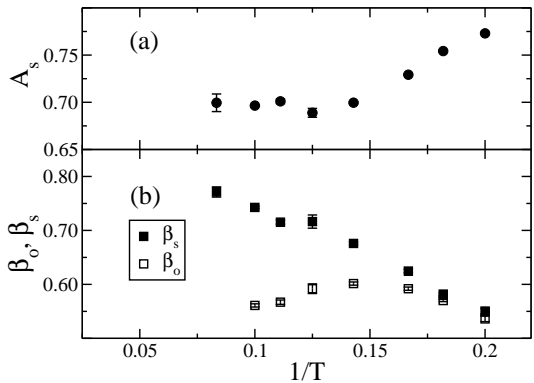}
\caption{\label{betaPlat} (a) The amplitude of the stretched exponential fit to the long-time decay of the self-intermediate
scattering function as a function of the inverse temperature. According to Ref.~\cite{Weysser:2010}, $A_s$ is 
a good estimate of the intermediate time plateau of $F_s(q;t)$.
(b) The stretching exponents $\beta_o$ and $\beta_s$ obtained from the stretched exponential fits to the long-time decay of the overlap functions and 
the self-intermediate scattering function, respectively, plotted versus the inverse temperature.}
\end{figure}

In Fig.~\ref{betaPlat} we show the temperature dependence of the amplitude of the stretched exponential fit
to the self-intermediate scattering function $A_s$ and the stretching exponents $\beta_o$ and $\beta_s$. The former quantity, $A_s$, is significant
in that it may be used to identify the mode-coupling regime. Specifically, the mode-coupling theory predicts that the self-intermediate
scattering function develops a plateau which extends to longer and longer times upon approaching the mode-coupling transition. The height
of this plateau is predicted to be temperature independent. While extracting the plateau height from data showed in Fig.~\ref{fs}
is somewhat delicate, Weysser \textit{et al.} \cite{Weysser:2010} showed that the amplitude of the stretched exponential fit
agrees with a value obtained from a more sophisticated procedure.

Fig.~\ref{betaPlat}(a) shows that the 
amplitude of the stretched exponential fit is approximately temperature independent down to $T=7$ and then starts increasing. 
This would suggest that the mode-coupling regime ends at this temperature. While the data showed in Fig.~\ref{betaPlat}(a) are
quite suggestive, we feel that this issue needs to be investigated further before making a strong claim that the mode-coupling regime ends at $T=7$. 
Therefore, in the fits described below we use two different temperature ranges, $12 \ge T \ge 7$ and $12 \ge T \ge 6$. We note that the upper limit of
the mode-coupling regime is determined by the fact that above $T=12$ a plateau in either the overlap function or the self-intermediate
scattering function cannot be determined.  The latter observation agrees reasonably well with value of the onset temperature of slow dynamics 
which has been found to be around $T = 13$ by Berthier \textit{et al.}\ \cite{Berthier:2012kd}.
Finally, we note that Fig.~\ref{betaPlat}(b) shows that at low temperatures the stretching exponents obtained
from fitting the final decays of $F_o(t)$ and $F_s(q;t)$ to stretched exponentials converge. We note that the overlap function can formally 
be obtained through an integral of the self-intermediate scattering function over all wave-vectors,
\begin{equation}\label{ovfs}
F_o(t) = \int \frac{d\mathbf{q}}{(2\pi)^3} f(q;a) F_s(q;t)
\end{equation}
where $f(k;a) = 4\pi a^2 j_1(ka)/k$ with $j_1$ denoting a spherical Bessel function of the first kind.
Eq. (\ref{ovfs}) together with the low-temperature convergence of the stretching exponents suggests that the characteristic relaxation time
$\tau_s$ obtained from the stretched exponential fit to the long-time decay of $F_s(q;t)$ and the stretching exponent $\beta_s$ 
are wave-vector independent, at least for the wave-vectors making the dominant contribution to the integral in Eq. (\ref{ovfs}). 
To investigate this further one needs to evaluate and analyze the self-intermediate scattering function for a range of
wave-vectors. 

\begin{figure}
\includegraphics[width=3.2in]{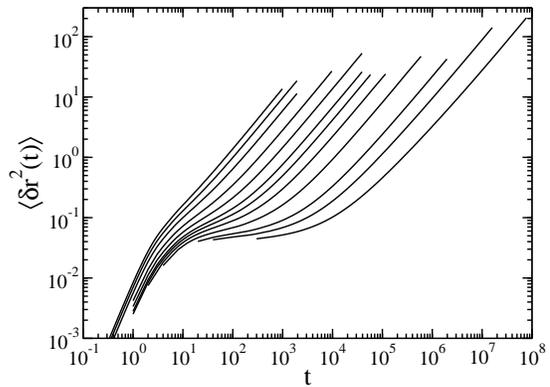}
\caption{\label{msd} The mean square displacement $\left< \delta r^2(t) \right>$ for $T=30$, 25, 20, 15, 12, 10, 9 8, 7, 6, 5.5, and
5, listed from left to right.}
\end{figure}

Another measure of single particle dynamics is the mean-square displacement
\begin{equation}
\left< \delta r^2(t) \right> = N^{-1} \left< \sum_n [\mathbf{r}_n(t) - \mathbf{r}_n(0)]^2 \right>,
\end{equation}
which is shown in Fig.~\ref{msd}. We clearly see a short time ballistic regime where $\left< \delta r^2(t) \right> = 3 T t^2$, 
and a diffusive regime where $\left< \delta r^2(t) \right> = 6 D t$. Furthermore, at low temperatures at intermediate 
times a plateau region emerges between the ballistic and diffusive regimes.  This plateau is associated with the cage effect where particles
are trapped inside a cage of neighboring particles. How the particles escape this cage
is one of the fundamental questions of the dynamics of glass-forming liquids. 

\begin{figure}
\includegraphics[width=3.2in]{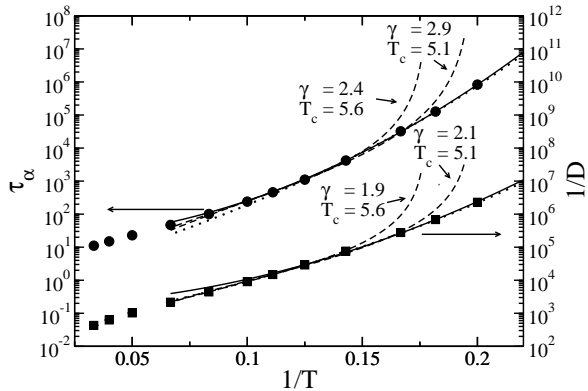}
\caption{\label{tauD}The relaxation time $\tau_\alpha$ and the inverse diffusion coefficient $D^{-1}$ 
versus $1/T$. The dashed lines are mode-coupling power law fits for two different fit ranges: $12 \ge T \ge 6$ and $12 \ge T \ge 7$. 
The solid lines are fits to $\tau_\alpha = \tau_0 e^{k_0 T^{-2}}$ and $D = D_0 e^{-d_0 T^{-2}}$ and the dotted lines are fits
to $\tau_\alpha = \tau_{v} e^{k_{v} (T-T_o)^{-1}}$ 
and $D = D_ve^{-d_{v} (T-T_o)^{-1}}$  where $T_o = 1.7$. The latter fits were performed using $T\le 8$ data.
}
\end{figure}

We determine the diffusion coefficient $D$ from the asymptotic, linear in time, dependence of the mean-square displacement,  
$D = \lim_{t\to\infty}\left< \delta r^2(t) \right>/(6t)$. In Fig.~\ref{tauD} we show the dependence of the $\alpha$-relaxation time
and the diffusion coefficient on the inverse temperature. Lines in this figure indicate fits discussed in the next two paragraphs.

According to the mode-coupling theory there should be one temperature $T_c$ at which the ergodicity is broken and one critical exponent $\gamma$.
To allow for better quality fits we followed Kob and Andersen \cite{Kob:1995,Kob:1995pr} and we fitted the $\alpha$-relaxation time data and 
the diffusion coefficient data to mode-coupling-like power laws $\tau_\alpha = a(T-T_c)^{-\gamma^\tau}$ and $D = b(T-T_c)^{\gamma^D}$.
In other words, we imposed one mode-coupling temperature $T_c$ common to 
both quantities but 
we allowed for different critical exponents. As mentioned above, we used two different temperature ranges. 
Fits using the $12 \ge T \ge 6$
result in $T_c = 5.1 \pm 0.1$, $\gamma^\tau = 2.9 \pm 0.1$ and $\gamma^D = 2.1 \pm 0.1$. Fits using $12 \ge T \ge 7$ 
result in $T_c = 5.6 \pm 0.1$, $\gamma^\tau = 2.4 \pm 0.1$ and $\gamma^D = 1.9 \pm 0.1$. 
All these fits are shown in Fig.~\ref{tauD} as dashed lines.

Also shown in Fig.~\ref{tauD} are two other commonly used fits. Solid lines show 
$\tau_\alpha = \tau_0 e^{k_0 T^{-2}}$ and $D = D_0 e^{-d_0 T^{-2}}$ and dotted lines show $\tau_\alpha = \tau_{v} e^{k_{v} (T-T_o)^{-1}}$ 
and $D = D_ve^{-d_{v} (T-T_o)^{-1}}$ . Both sets of fits were performed for $T \le 8$. 
Note that the latter set of fits imposes the same Vogel-Fulcher temperature
$T_o$ for both the $\alpha$-relaxation time and the diffusion coefficient. This set of fits results in $T_o=1.7$. 
We shall emphasize that either $e^{B T^{-2}}$ or $e^{A (T-T_o)^{-1}}$ fits describe very well the low temperature data.
Thus, our simulational results cannot exclude either a $T=0$ or a finite temperature divergence of the $\alpha$-relaxation time. 

We note that the mode-coupling temperature obtained in this work from fits using 
the range of temperatures $12 \ge T \ge 6$, $T_c=5.1 \pm 0.1$, is slightly smaller but within error of
the $T_c=5.2$ reported by Kob \textit{et al.} \cite{Kob:2011cd}. The latter value was obtained from fitting the $\alpha$-relaxation times obtained from
two separate self-intermediate scattering functions involving small and large particles. In Appendix A we show that the fit of the $\alpha$-relaxation 
obtained from the overlap function involving all particles results in a value comparable to that of Kob \textit{et al.} In the same Appendix we
also discuss more elaborate fitting procedures using relaxation times and diffusion coefficients pertaining to small and large particles.
The discussion above and in Appendix A suggests that instead of the single mode-coupling temperature $T_c$ we should introduce a range
of mode-coupling temperatures or a $T_c$ range. We propose to identify the temperature range $5.6-5.1$ as the $T_c$ range for the system studied 
in this work and we will indicate this range in the figures rather than showing a single mode-coupling temperature.  

The above analysis shows that 
our results are very similar to what is experimentally observed for super-cooled liquids \cite{Berthier:2011hs}. There is a range
of temperatures where mode-coupling like power laws provide a reasonable description of the data. While the mode-coupling regime is reasonably
well defined, small changes in the range of temperatures used for fitting result in somewhat large changes of the mode-coupling temperature. 
Importantly, neither the relaxation time diverges nor the diffusion coefficient vanishes 
at the $T_c$ inferred from the fits, and any mode-coupling-like transition is
avoided. Furthermore, the exponents associated with the power-law fits for $\tau_\alpha$ and
for $D$ differ from each other, which is contrary to the predictions of mode-coupling theory. 

\begin{figure}
\includegraphics[width=3.2in]{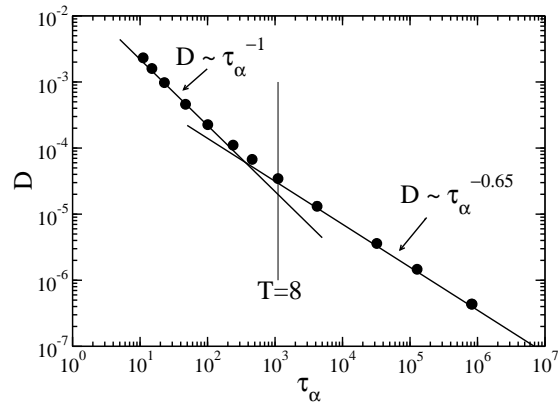}
\caption{\label{stokes}Stokes-Einstein decoupling. For $T > 12$, $D \sim \tau_\alpha^{-1}$,
but for $T < 8$ we find that the fractional Stokes-Einstein relationship $D \sim \tau_\alpha^{-0.65}$
provides an accurate description of the data. The thin solid vertical line indicates the relaxation time corresponding to 
temperature $T=8$.}
\end{figure}

The difference
in the mode-coupling-like power laws' exponents is not surprising when one realizes that the Stokes-Einstein relation is violated, see Fig. \ref{stokes}. 
For high temperatures we find that Stokes-Einstein relation is obeyed, $D \sim \tau_\alpha^{-1}$
till about $T=12$.
In contrast, for the lowest temperatures we find that a fractional Stokes-Einstein relation works well, 
$D \sim \tau_\alpha^{-0.65}$. This fractional Stokes-Einstein relation emerges for $T<8$, which is 
inside both temperature ranges used for the mode-coupling-like power law fits. Thus, the exponent for the fit to $\tau_\alpha$
and $D$ would not expected to be the same.  Also, we note that there is a temperature range, $12 > T > 8$, where 
the standard Stokes-Einstein relation is violated, and the fractional Stokes-Einstein behavior has yet to emerge. 

In summary, for temperatures lower than what is well described by the power laws, a new behavior emerges
which is well described by an exponential divergence of the relaxation time and 
an exponential vanishing of the diffusion coefficient. The nature of this divergence
is under debate since different functional forms describe many different glass
formers well (see \textit{e.g.}\ Refs.~\cite{Berthier:2011hs,Elmatad:2009ec} and
references therein). In the following sections, we will explore the relationship between the liquid's dynamics
and dynamic heterogeneities. 

\section{Dynamic Susceptibility and Correlation Length}\label{sec:length}
In the previous section we examined different aspects of the single particle dynamics within the 
model glass-forming liquid studied here. We observed a change in many aspects of the dynamics as
the temperature decreases. In particular, the final  
relaxation at low temperatures is broadened and is 
well described by a stretched exponential. Stretched exponential relaxation can result from different populations of 
particles where some populations relax much faster or much slower than would expected on the basis of a Gaussian distribution. 
To look for these subsets of particles we examine the self-van Hove correlation function,
$G_s(\delta r;t) = \left< \delta r - [\mathbf{r}_n(t) - \mathbf{r}_n(0)] \right>$. Since short time ballistic and long 
time diffusive motion both result in Gaussian distributions of $G_s(\delta r;t)$, the existence of fast and slow
sub-populations may be found by comparing $G_s(\delta r;t)$ to a Gaussian distribution
with the mean-square displacement equal to that calculated from the simulation. 
Following a procedure suggested previously \cite{Puertas:2004,Flenner:2005es}, 
we present here the probability distributions of the logarithm of single-particle
displacements which is easily obtained from the self-van Hove correlation function,
$P[\log_{10}(\delta r);t] = \ln(10) 4 \pi \delta r^3 G_s(\delta r,t)$. The advantage of looking at 
this probability distribution is that if the van Hove function is Gaussian, the shape of $P[\log_{10}(\delta r);t]$
is independent of time. 

\begin{figure}
\includegraphics[width=3.2in]{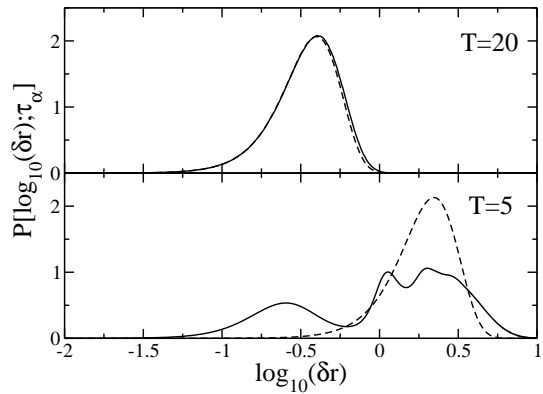}
\caption{\label{psmall}Comparison of $P[\log_{10}(\delta r);\tau_\alpha]$ for the smaller particles at 
$T=20$ and $T=5$ (solid lines) with what would be
expected from a Gaussian distribution (dashed lines) with 
the same mean-square displacement.}
\end{figure}

\begin{figure}
\includegraphics[width=3.2in]{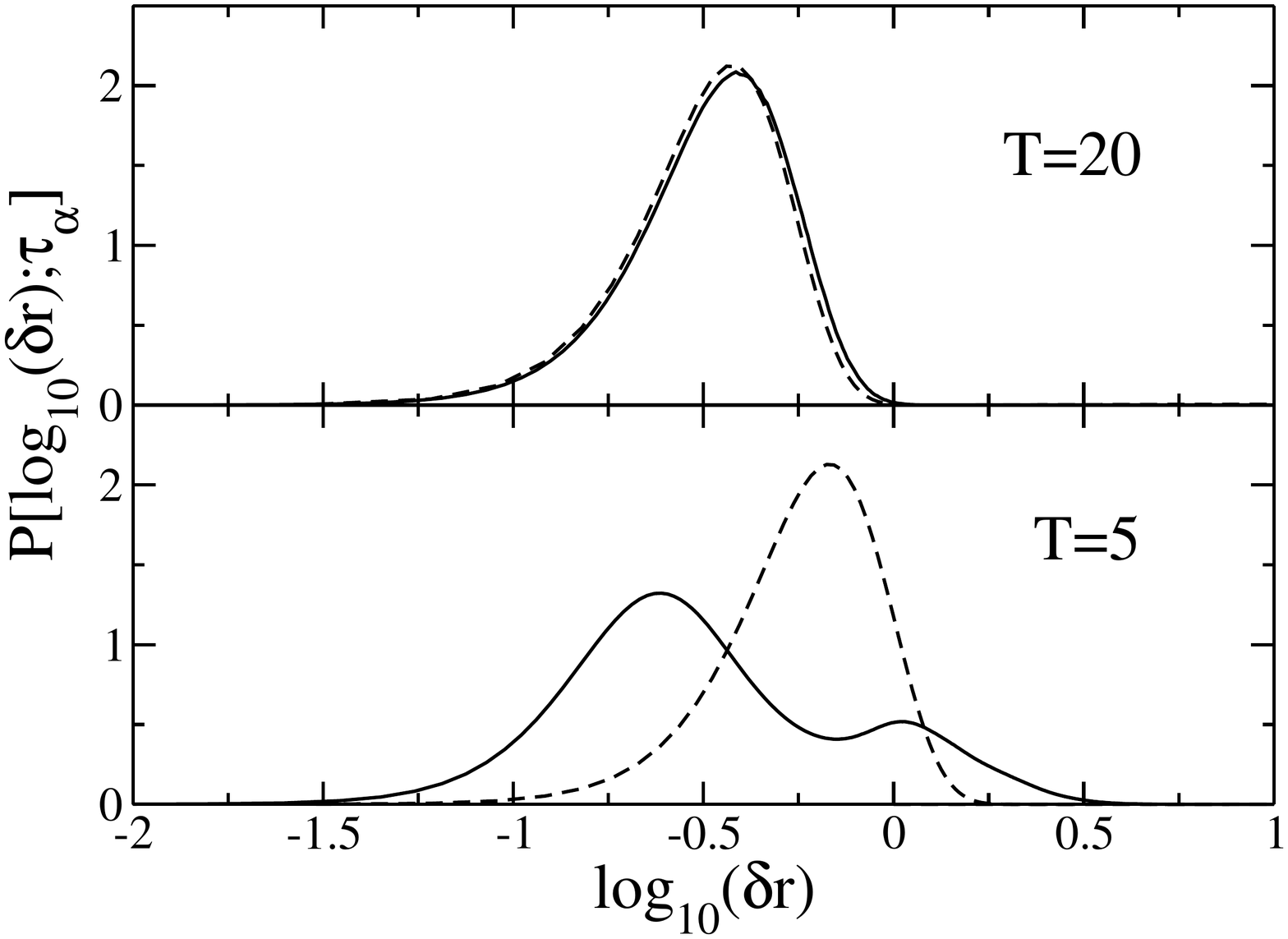}
\caption{\label{plarge}Comparison of $P[\log_{10}(\delta r);\tau_\alpha]$ for the larger particles at
$T=20$ and $T=5$ (solid lines) with what would be
expected from a Gaussian distribution (dashed lines) with 
the same mean-square displacement.}
\end{figure}

Shown in Figs.~\ref{psmall} and \ref{plarge} are the calculated probabilities of the 
logarithm of single particle displacements at the $\alpha$-relaxation time for $T=20$ and $T=5$
compared to what would be obtained from a Gaussian van Hove function with the same mean-square displacement $\left< \delta r^{2}(\tau_\alpha) \right>$,
calculated using the smaller and larger particles. The Gaussian distribution 
describes the results well for $T=20$, before the onset of supercooling. In contrast, at 
$T=5$, which is below the mode-coupling temperature, we observe several peaks, which suggests
an activated, hopping-like relaxation mechanism. 
While Figs.~\ref{psmall}-\ref{plarge} clearly indicate the existence of sub-populations of slow and fast particles, they cannot
reveal spatial correlations of these sub-populations of particles. 

\begin{figure}
\includegraphics[width=3.2in]{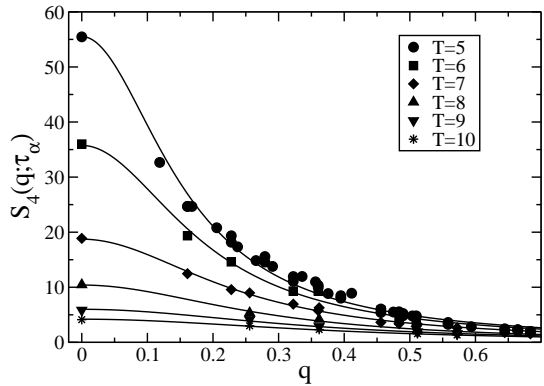}
\caption{\label{fig:s4over}The four-point structure factor at some representative temperatures. The solid lines
are fits to the Ornstein-Zernicke equation $\chi_4(\tau_\alpha)/[1+(\xi(\tau_\alpha) q)^2]$ for 
$q < 1.5/\xi_4(\tau_\alpha)$.}
\end{figure}

To examine the spatial correlations of slow particles we followed previous work \cite{Lacevic:2003vq,Lacevic:2002bi,Flenner:2011ht,Flenner2010prl} and 
examined the four-point structure factor
\begin{eqnarray}
S_4(q;t) &=& \frac{1}{N} \left< \sum_{n,m} w_n(t) w_m(t) e^{-i \mathbf{q} \cdot [\mathbf{r}_n(0) - \mathbf{r}_m(0)]} \right>\nonumber \\
&& - \frac{1}{N} \left| \left<  \sum_n w_n(t) e^{-i\mathbf{q} \cdot \mathbf{r}_n(0)} \right> \right|^2.
\end{eqnarray} 
$S_4(q;t)$ is the structure factor calculated using the particles that have moved less than a distance
$a$ during a time $t$. It measures correlations between microscopic overlap functions pertaining to individual particles. 
For $t=0$, $S_4(q;t=0)$ is equal to the static structure factor. At later times 
there is an enhancement of the low $q$ values which indicates that the slow particles, \textit{i.e.}\ the
particles that have moved less than $a$ over the time $t$, form clusters (see Fig. \ref{fig:s4over} for 
the four-point structure factor at the $\alpha$-relaxation time). To examine
these clusters we study the dynamic susceptibility $\chi_4(t) = \lim_{q\rightarrow0} S_4(q;t)$,
which is related to the number of dynamically correlated particles, and
the dynamic correlation length $\xi_4(t)$, which is a measure of the size of the slow regions.

To aid in fits which we use to determine the correlation length the  we calculated
\begin{equation}\label{chi4def}
\chi_4(t) = N^{-1}\left( \left< \sum_{m,n} w_n(t) w_m(t) \right> - \left< \sum_n w_n(t) \right>^2 \right)
\end{equation}
from simulations and then used it at $t=\tau_\alpha$ as a fit point at $S_4(q=0;\tau_\alpha)$. 
We note that $\chi_4(t)$ as given by Eq. (\ref{chi4def}) is equal to the small wave-vector limit of the four-point structure factor $\lim_{q\to 0}S_4(q;t)$
only if the average in Eq. (\ref{chi4def}) is performed in the ensemble that does not suppress global fluctuations 
\cite{Lebowitz:1967vs, Berthier:2007jcp1, Berthier:2007jcp2}. Thus, one 
can not directly calculate it in the $NVE$ and $NVT$ simulations since energy and/or particle number fluctuations are suppressed 
in these simulational ensembles. However,
$\chi_4(t)$ can be calculated by taking into account these suppressed fluctuations 
as described in Refs.~\cite{Flenner2010prl,Flenner:2011ht,Lebowitz:1967vs}. 
Details of this method for the system studied here are given in Appendix B. For simplicity and to aid in the following 
discussion we introduce the following notation for a contribution due to energy fluctuations $\chi_4(t)|_{\delta T} = k_B T^2 \chi_T^2(t)/c_v$ 
where $k_B$ is Boltzmann's constant, $c_v$ is the constant volume specific heat per particle, 
and $\chi_T(t) = \partial F_o(t)/\partial T$. In addition, we introduce the following notation for a contribution
due to particle numbers fluctuations $\chi_4(t)|_{\delta N}$. The expression for the latter quantity is rather long and it is given in Appendix B.
As discussed before \cite{Flenner2010prl,Flenner:2011ht,Lebowitz:1967vs,Berthier:2007jcp1,Berthier:2007jcp2}
\begin{equation}\label{chi4}
\chi_4(t) = \chi_4(t)|_{NVE} + \chi_4(t)|_{\delta T} + \chi_4(t)|_{\delta N}
\end{equation}
where $\chi_4(t)|_{NVE}$ is $\chi_4(t)$ given by Eq. (\ref{chi4def}) calculated in an $NVE$ simulational ensemble. Likewise we denote
$\chi_4(t)|_{NVT}$ for $\chi_4(t)$ calculated in the $NVT$ simulational ensemble. 
At $T=5.5$ we checked that 
\begin{equation}\label{chi4T}
\chi_4(\tau_\alpha)|_{NVT} = \chi_4(\tau_\alpha)|_{NVE} + \chi_4(\tau_\alpha)|_{\delta T}
\end{equation}
by calculating all terms in Eq. (\ref{chi4T}) separately. 
For $T=5$, due to energy drift in the $NVE$ simulations, caused by the very long simulation
time (over 4 billion time steps) required, we used $\chi_4(\tau_\alpha)|_{NVT}$
for the sum $\chi_4(\tau_\alpha)|_{NVE} + \chi_4(\tau_\alpha)|_{\delta T} $. It is important to note that poor choices of 
simulation parameters can result in 
$\chi_4(\tau_\alpha)|_{NVT} \ne \chi_4(\tau_\alpha)|_{NVE} + \chi_4(\tau_\alpha)|_{\delta T}$ due to too strong or 
too weak of a coupling to the temperature bath; see Appendix B for more details.

Shown in Fig.~\ref{terms} are $\chi_4(\tau_\alpha)|_{NVE}$ (open circles), $\chi_4(\tau_\alpha)|_{\delta T}$ 
(squares) and $\chi_4(\tau_\alpha)|_{\delta N}$ (triangles) for $T \le 15$. At high temperatures, $\chi_4(\tau_\alpha)|_{NVE}$
is the largest term. For $T < 10$, $\chi_4(\tau_\alpha)|_{NVE}$ grows slower than $\chi_4(\tau_\alpha)|_{\delta N}$ and for $T \le 6$ the latter
term becomes the largest contribution to $\chi_4(\tau_\alpha)$. For $T < 8$ both 
$\chi_4(\tau_\alpha)|_{\delta N}$ and $\chi_4(\tau_\alpha)|_{\delta T}$ grow as $T^{-4}$ with decreasing temperature. 
We note that $T=8$ is the temperature at which the fractional Stokes-Einstein behavior emerges for this system, Fig.~\ref{stokes}.
At $T=5$ we could not perform a long enough
$NVE$ simulation without significant energy drift, and a calculation $\chi_4(\tau_\alpha)|_{NVE}$ for short 
simulations results in a very large uncertainty. However, by looking at 
$\chi_4(\tau_\alpha)|_{NVT} - \chi_4(\tau_\alpha)|_{\delta T}$ it appears that $\chi_4(\tau_\alpha)|_{NVE}$ at 
$T=5$ decreases. If the above discussed trends continue to lower temperatures, then  
$\chi_4(\tau_\alpha)$ should also grow as $T^{-4}$ with decreasing temperature. 
\begin{figure}
\includegraphics[width=3.2in]{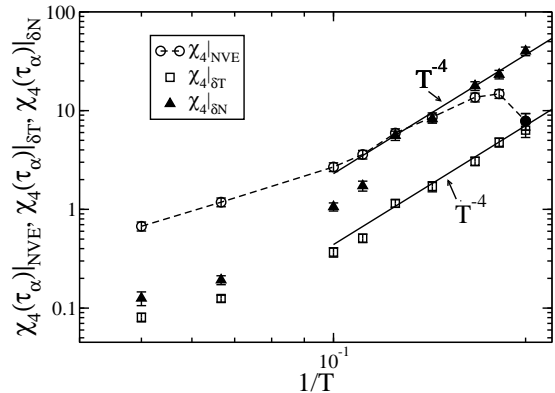}
\caption{\label{terms}The different terms which represent the main contributions to the ensemble independent
susceptibility $\chi_4(\tau_\alpha)$. The circles are $\chi_4(\tau_\alpha)|_{NVE}$ and are directly calculated in 
an $NVE$ simulation (open circles) except for at $T=5$ (filled circle) where we used 
$\chi_4(\tau_\alpha)|_{NVE} = \chi_4(\tau_\alpha)|_{NVT} - \chi_4(\tau_\alpha)|_{\delta T}$. The dashed line through the circles serves to guide the eyes.}
\end{figure} 

Having $S_4(q;\tau_\alpha)$ for $q>0$ from direct simulations and using $\chi_4(\tau_\alpha)$ from the 
above described procedure as $S_4(q=0;\tau_\alpha)$, 
we now fit the four-point structure factor to an Ornstein-Zernicke function 
\begin{equation} 
\frac{\chi_4(\tau_\alpha)}{1+(\xi_4(\tau_\alpha) q)^2}.
\end{equation}
We determined previously \cite{Flenner:2011ht,Flenner2010prl} that these fits are robust if the range of wave-vectors used in
the fits is restricted to $q < 1.5/\xi_4(\tau_\alpha)$. The Ornstein-Zernicke fits are showed as solid lines in Fig.~\ref{fig:s4over}.
We note that the fits result in values of $\chi_4(\tau_\alpha)$ that are slightly different from those determined by
the procedure described above. These differences are seen in Fig.~\ref{fig:s4over} as differences between data points at $q=0$ and
the positions of the solid lines at $q=0$. In the remainder of the paper, we will 
plot and discuss $\chi_4(\tau_\alpha)$ as determined from the Ornstein-Zernicke fits.

\begin{figure}
\includegraphics[width=3.2in]{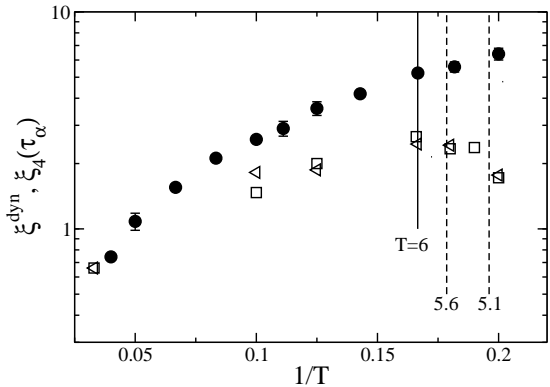}
\caption{\label{xiT} Comparison of different dynamic correlation lengths as a function of  $1/T$. 
The circles are $\xi_4(\tau_\alpha)$ calculated in this work, the left triangles are $\xi_c^{\mathrm{dyn}}$
and the squares are $\xi_s^{\mathrm{dyn}}$. The latter two lengths are the dynamic point-to-set lengths
described in Ref.~\cite{Kob:2011cd}. The solid vertical line shows the maximum of 
$\xi^{\mathrm{dyn}}$ and the dashed lines gives the range of mode-coupling temperatures
that are consistent with mode-coupling like power law fits to $\tau_\alpha$ and $D$.}
\end{figure}

In Fig. \ref{xiT} we compare the temperature dependence of the four-point dynamic correlation length determined in this work and the point-to-set dynamic 
lengths of Kob \textit{et al.} These lengths have a similar magnitude at high temperatures. The four-point length $\xi_4$ increases 
with decreasing temperature faster than either of the two lengths determined by Kob \textit{et al.} Most importantly, $\xi_4$ increases
with decreasing temperature for all temperatures examined in this work whereas the lengths determined by Kob \textit{et al.} peak around $T=6$
and then decrease with decreasing temperature. We note that the latter behavior correlates with the non-monotonic temperature 
dependence of the finite size effects investigated by Berthier \textit{et al.} \cite{Berthier:2012kd}. 
Curiously, our four-point dynamic correlation lengths are significantly longer
than the point-to-set \emph{static} correlation lengths
determined by Kob \textit{et al.} This contrast between four-point and static point-to-set correlation lengths is also present in
other systems (compare, \textit{e.g.} dynamic lengths determined in Ref. \cite{Flenner2009} and static lengths of Ref. \cite{Hocky2012}). 

\begin{figure}
\includegraphics[width=3.2in]{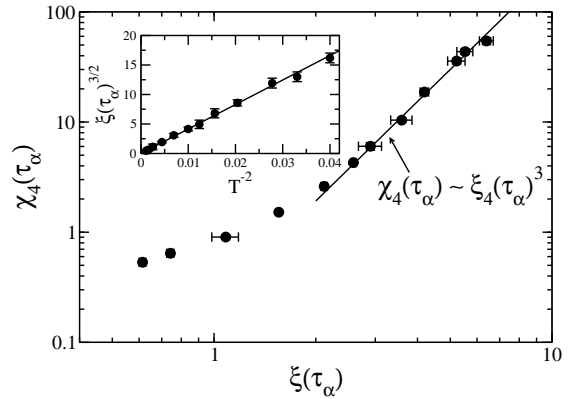}
\caption{\label{chixi}The relationship between $\chi_4(\tau_\alpha)$ and $\xi_4(\tau_\alpha)$ 
where $\chi_4(\tau_\alpha)$ and $\xi_4(\tau_\alpha)$ are both obtained from the
Ornstein-Zernicke fits to $S_4(q;\tau_\alpha)$. The
straight line is a fit to $\chi_4(\tau_\alpha) \sim \xi_4(\tau_\alpha)^{3}$, and describes the data well
for $T<9$. The inset is $\xi_4(\tau_\alpha)^{3/2}$ versus $T^{-2}$, which describes the data well over 
the whole temperature range.}
\end{figure}

Next we study the relationship between $\chi_4(\tau_\alpha)$ and $\xi_4(\tau_\alpha)$, which
is shown in Fig.~\ref{chixi}. As in previous work \cite{Flenner2010prl,Flenner:2011ht}, 
we find that $\chi_4(\tau_\alpha) \sim \xi_4(\tau_\alpha)^3$
when $\xi_4(\tau_\alpha)$ is greater than approximately 3, which occurs at 
a temperature of approximately 9. This result suggests compact domains of slow particles.
We do, however, caution on obtaining a correlation length from $\xi_4(t) \sim [\chi_4(t)]^{1/3}$.
It was observed in previous work, Ref.~\cite{Flenner:2011ht}, that $\xi_4(t)$ reaches a
plateau and remains constant as a function of time during a period when $\chi_4(t)$ decreases. Thus, to compare 
$\xi_4(t)$ at different temperatures, different times, under confinement, etc.\ one should not use the relationship 
$\xi_4(t) \sim  [\chi_4(t)]^{1/3}$. 

We now examine the relationship between $\tau_\alpha$ and $\xi_4(\tau_\alpha)$. Previously 
we observed for a hard-sphere glass-forming system that $\tau_\alpha \sim e^{k_2 \xi_4}$
provided a good description of the results over the full range 
of volume fractions studied in Refs.\ \cite{Flenner:2011ht,Flenner2010prl}. 

\begin{figure}
\includegraphics[width=3.2in]{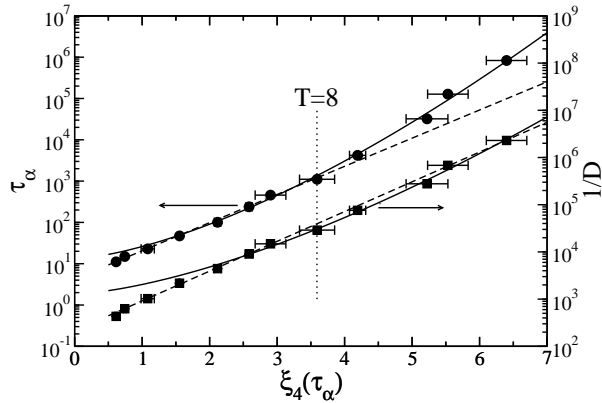}
\caption{\label{tauDxi} The $\alpha$ relaxation time $\tau_\alpha$ (circles, left axis) and $D^{-1}$ 
(squares, right axis) as a function 
of $\xi_4(\tau_\alpha)$. The solid lines are fits to $\tau_\alpha = \tau_1 e^{k_1 \xi_4^{3/2}}$ and
$D = D_1 e^{-d_1 \xi_4^{3/2}}$, while the dashed lines are fits to $\tau_\alpha = \tau_2 e^{k_2 \xi_4}$ 
and $D = D_2 e^{-d_2 \xi_4}$.}
\end{figure}

In Fig.~\ref{tauDxi} we show $\tau_\alpha$ versus $\xi_4(\tau_\alpha)$ for the harmonic spheres. 
For $T \ge 8$ we find that $\tau_\alpha = \tau_2 e^{k_2 \xi_4}$,
but for $T \le 8$ we find a faster growth of the $\alpha$-relaxation time with the dynamic 
correlation length, $\tau_\alpha = \tau_1 e^{k_1 \xi_4^{3/2}}$. We note that
the Random First Order theory \cite{Wolynes} predicts that $\tau_\alpha \sim e^{\xi_s^{3/2}}$
where $\xi_s$ is a time-independent static correlation length characterizing the size
of correlated regions, and this functional form is consistent with our results.

Exponential correlations shown in Fig. \ref{tauDxi} combined with 
power law relations between the diffusion coefficient and the $\alpha$-relaxation
time, $D \sim \tau_\alpha^{-1}$ and  $D \sim \tau_\alpha^{-0.65}$ (see Fig. \ref{stokes}),
imply similar exponential correlations between the diffusion coefficient and the dynamic correlation length.
In Fig.~\ref{tauDxi} we also show $1/D$ versus $\xi_4(\tau_\alpha)$ for the harmonic spheres. 
We find that $D = D_2 e^{-d_2 \xi_4}$ fits the data well for all temperatures except $T=7$.
However, the $T \le 8$ data are also compatible with $D = D_1 e^{-d_1 \xi_4^{3/2}}$. 
Simulations for somewhat lower temperatures, which do not seem feasible at present, 
are needed to shed some additional light at the relation between the diffusion coefficient and
the dynamic correlation length.

We note the internal consistency of our low temperature correlations. If the observed trends continue
then for low temperatures $\chi_4(\tau_\alpha) \approx \chi_4(\tau_\alpha)|_{\delta T} + \chi_4(\tau_\alpha)|_{\delta N}$ 
and $\chi_4(\tau_\alpha) \sim T^{-4}$. The last dependence, taken together with
the correlation $\chi_4(\tau_\alpha) \sim \xi^3$, implies that at low temperatures $\xi^{3/2} \sim T^{-2}$.
We find that $\xi^{3/2} \sim T^{-2}$ describes the data well for the whole temperature range studied in 
this work even for temperatures where $\chi_4(\tau_\alpha)$ does not increase as
$T^{-4}$; see the inset to Fig.~\ref{chixi}.
Since we find that $\tau_\alpha \sim e^{k_0 T^{-2}}$ provides a good description of the data for $T\le 9$ (see Fig. \ref{tauD}), one would 
expect that $\tau_\alpha \sim e^{k_1 \xi^{3/2}}$ to also provide a good description for these temperatures.  

Our data is also consistent with a divergence of $\xi_4 \sim (T-T_0)^{-2/3}$, but to obtain
a $T_0$ consistent with the fits to $\tau_\alpha$ and $D$ we must introduce an offset 
and fit \textit{e.g.} $\xi_4 = (a_\xi + b_\xi/(T-T_0))^{2/3}$.  A similar sort of offset is also needed for
fits to $\chi_4(\tau_\alpha)$. 

In summary, in this section we examined the relationships between $\xi_4(\tau_\alpha)$,
$\tau_\alpha$ and $D$. Our results are consistent with $\tau_\alpha = \tau_2 e^{k_2 \xi_4}$ and
$D = D_2e^{-d_2 \xi_4}$ for $T > 8$ and then a crossover to $\tau_\alpha = \tau_1 e^{k_1 \xi_4^{3/2}}$ and
$D = D_1 e^{-d_1 \xi_4^{3/2}}$ for $T \le 8$. We also found that this relationship is consistent with the 
temperature dependence of the correction terms to $\chi_4(\tau_\alpha)$ and the observed fractional
Stokes-Einstein relation examined in the last section. 

\section{Conclusions}\label{sec:summary}
We examined dynamic heterogeneities above and below the mode-coupling temperature and found
evidence of a dynamic crossover associated with the dynamic heterogeneities. To identify the mode-coupling
temperature we fit the $\alpha$-relaxation time and the diffusion coefficient to mode-coupling like
power laws. We obtained a slightly different 
mode-coupling temperatures $T_c$ depending on the range of temperatures used for fitting. 
We found that mode-coupling like power law fits to $\tau_\alpha$ and and $D$ 
give different exponents for this system. This is consistent with the onset of the Stokes-Einstein relation  
violation inside the mode-coupling regime. Specifically, we found that $D \sim \tau_\alpha^{-0.65}$ beginning around $T=8$. 
However, an exponential divergence of $\tau_\alpha$ and $D$ with the same form 
for both $\tau_\alpha$ and $D$ is consistent with a fractional Stokes-Einstein relation, and
we find that two commonly used forms $e^{B T^{-2}}$ and $e^{A (T-T_0)^{-1}}$ are both consistent with our data.  

We examined and characterized dynamic heterogeneities from $T=20$ down to $T=5$, which includes 
temperatures above the onset of slow dynamics, $T\approx 13$ \cite{Berthier:2012kd}, and below the mode-coupling temperature range, 
$T_c \approx 5.6-5.1$. We calculated the four-point susceptibility $\chi_4(\tau_\alpha)$ and
the correlation length $\xi_4(\tau_\alpha)$ by fitting the four-point structure factor $S_4(q;\tau_\alpha)$
to an Ornstein-Zernicke function.
Through the relationships between $D$, $\tau_\alpha$, and $\xi_4$ we found a dynamic 
crossover that had remained hidden in a previous hard sphere simulation \cite{Flenner:2011ht,Flenner2010prl} due to a smaller
range of relaxation times.
We found that $\tau_\alpha \sim e^{k_2 \xi_4}$ for
$T > 8$, which includes the first decade of slowing down after the onset of slow dynamics at $T\approx 13$, then 
$\tau_\alpha \sim e^{k_1 \xi_4^{3/2}}$ for $T<8$, where $T=8$ is the same temperature
where we find the onset of the fractional Stokes-Einstein relationship. In contrast, we found
that $D \sim e^{-d_2 \xi_4}$ fits our data well over the whole temperature range. We note that to be 
consistent with the fractional Stokes-Einstein relationship one would expect  
$D \sim e^{-d_1 \xi_4^{3/2}}$ for $T<8$ and we found that this is also consistent with our data. Simulations at lower temperatures 
are needed to examine the relationship between $D$ and $\xi$.

For $T<8$ there is some interesting behavior of different contributions to the
four-point dynamic susceptibility $\chi_4(\tau_\alpha)$. In our simulations we directly
calculate the constant volume and energy contribution $\chi_4(\tau_\alpha)|_{NVE}$
or the constant volume and temperature contribution $\chi_4(\tau_\alpha)|_{NVT}$, and
to obtain the $\lim_{q\rightarrow0}$ of $S_4(q;\tau_\alpha)$ we calculate 
fluctuations suppressed in the $NVE$ or $NVT$ simulations. These correction terms
are related to the derivatives of the two point correlation function 
$\chi_x(t) = \partial F_o(t)/\partial x$.  We find that $\chi_\rho^2$ and $\chi_T^2$ both
grow as $T^{-4}$ starting around $T=8$; again around the same temperature where
the fractional Stokes-Einstein relationship emerges.  

We see evidence that $\chi_4(\tau_\alpha)|_{NVE}$ levels off around $T=6$ and
decreases around $T=5$. The similarity of the temperature dependence
of $\chi_4(\tau_\alpha)|_{NVE}$ and the point-to-set dynamic correlation length $\xi^{\mathrm{dyn}}$ 
studied by Kob \textit{et al.}\ \cite{Kob:2011cd}
is striking. Furthermore, 
this change in behavior of $\chi_4(\tau_\alpha)|_{NVE}$ and $\xi^{\mathrm{dyn}}$ occurs in a temperature
range where Berthier \textit{et al.}\ \cite{Berthier:2012kd} noticed a change in the finite size effects. This change in the finite
size effects is correlated with an avoided mode-coupling transition. 
It occurs in the temperature range where reasonable mode-coupling fits predict the
mode-coupling temperature in the binary harmonic spheres studied in this, Kob's, and 
Berthier's work.  More work seems to be needed to understand the complex dynamics around
the mode-coupling temperature range. 
\section*{Acknowledgments}
We gratefully acknowledge the support of NSF Grant CHE 0909676. This research utilized the CSU ISTeC Cray HPC System supported by NSF 
Grant CNS 0923386.


\section*{Appendix A}\label{sec:appendixA}

As discussed in Sec. \ref{sec:dynamics}, our results for the mode-coupling temperature depend on the range of temperatures used for
fitting. The fit using the temperature range $12 \ge T \ge 6$ results in $T_c = 5.1\pm 0.1$ which is slightly smaller but within error bars of 
$T_c=5.2$ of Kob \textit{et al.} \cite{Kob:2011cd}. Here we examine some alternative fits that can be used for the $\alpha$-relaxation time and diffusion
coefficient data obtained from our simulations. In particular, we show that in addition to the dependence of the mode-coupling temperature
on the range of temperatures used for fitting, there is also a somewhat unexpected dependence on the quantity being fit. 

\begin{figure}
\includegraphics[width=3.2in]{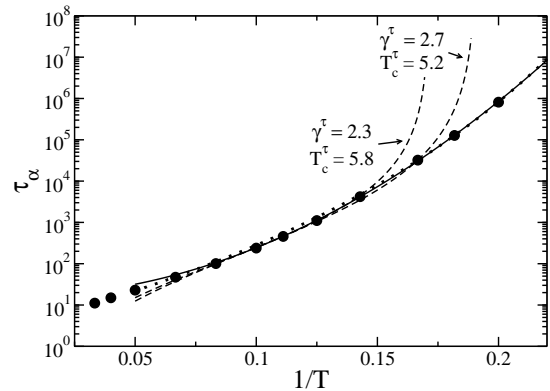}
\caption{\label{alpha} The $\alpha$-relaxation time as a function of the inverse temperature. The dashed lines are mode-coupling-like 
fits to the function $\tau_\alpha = a(T-T_c^\tau)^{-\gamma^\tau}$ where the parameters of the fits are given in the figure. 
Two other fits are shown in the figure, a solid
line which is a fit to $\tau_\alpha = \tau_0 e^{k_{0} T^{-2}}$ for $T\le8$, and a dotted line which is a fit
to $\tau_\alpha = \tau_{v} e^{k_{v} (T-T_o^\tau)^{-1}}$ with $T_o^\tau = 2.2$. For the latter fits we used $T\le 8$ data. 
At the lowest temperatures the solid and dotted line nearly overlap.} 
\end{figure}

In Fig.~\ref{alpha} we show the temperature dependence of the $\alpha$-relaxation time. The dashed lines are fits to 
a mode-coupling like power law $\tau_\alpha = a(T-T_c^\tau)^{-\gamma^\tau}$. A fit using the temperature range $12 \ge T \ge 6$ results in 
$T_c^\tau = 5.2 \pm 0.1$ and $\gamma^\tau = 2.7 \pm 0.1$, but a fit using the range $12 \ge T \ge 7$ results in $T_c^\tau = 5.8 \pm 0.1$
and $\gamma^\tau = 2.3 \pm 0.1$. Here the superscript $\tau$ denotes the mode-coupling temperature obtained from the $\alpha$-relaxation time fits. 
We note that the former result agrees very well with the result obtained by  Kob \textit{et al.} \cite{Kob:1994, Kob:1995} by fitting the $\alpha$-relaxation
times obtained from the self-intermediate scattering functions for small and large particles. 

Also shown in Fig.~\ref{alpha} are two other commonly used fits, $\tau_\alpha = \tau_0 e^{k_0 T^{-2}}$ (solid line)
and $\tau_\alpha = \tau_{v} e^{k_{v} (T-T_o^\tau)^{-1}}$ (dotted line). Both fits were performed for $T \le 8$, and both
fit the data very well for this temperature range. The latter fit results in $T_o^\tau = 2.2$ where, again, 
the superscript $\tau$ denotes the Vogel-Fulcher temperature obtained from the $\alpha$-relaxation time fit.

\begin{figure}
\includegraphics[width=3.2in]{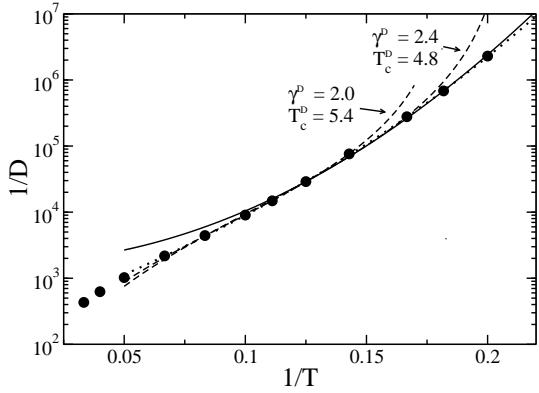}
\caption{\label{diff} 
The inverse of the diffusion coefficient as a function of the inverse temperature. The dashed lines are mode-coupling-like 
fits $D \sim (T-T_c^D)^{\gamma^D}$ with the values of the fit parameters given in the figure. 
Two other fits are shown in the figure, a solid
line which is a fit to $D = D_0 e^{-d_{0} T^{-2}}$ and a dotted line which is a fit
to $D = D_{v} e^{-d_{v} (T-T_o^D)^{-1}}$ with $T_o^D = 1.6$. For the latter fits we used $T\le 8$ data.}
\end{figure}

In Fig.~\ref{diff} we show the temperature dependence of the diffusion coefficient. The dashed lines are the mode-coupling-like 
fits $D = b (T-T_c^D)^{\gamma^D}$ for the same ranges of temperatures as used in the analysis of the $\alpha$-relaxation time. Fit using the 
range $12 \ge T \ge 6$
results in $T_c^D = 4.8 \pm 0.1$ and $\gamma^D = 2.4 \pm 0.1$, but a fit for $12 \ge T \ge 7$ results in $T_c^D = 5.4 \pm 0.1$
and $\gamma^D =  2.0 \pm 0.1 $. Here the superscript $D$ denotes the mode-coupling temperature obtained from the diffusion coefficient fits. 

We note that unlike what was found in the pioneering analysis of a binary Lennard-Jones system by Kob and Andersen \cite{Kob:1994}, 
we found that the mode-coupling temperatures determined from fits of the $\alpha$-relaxation time and the diffusion coefficient
are somewhat different. However, we also note that a visual comparison of Fig. \ref{tauD} with Figs. \ref{alpha} and \ref{diff}
shows that  in the temperature ranges used for fitting 
the fits imposing the same mode-coupling temperature for the $\alpha$-relaxation time and the diffusion coefficient 
seem as good as the fits allowing for different mode-coupling temperatures. 

Also shown in Fig.~\ref{diff} are two other commonly used fits, $D = D_0 e^{-d_0 T^{-2}}$ (solid line)
and $D = D_{v} e^{-d_{v} (T-T_o^D)^{-1}}$ (dotted line). Again, both fits were performed for $T \le 8$, and both
fit the data very well for this temperature range. The latter fit results in $T_o^D = 1.5$ which, again, is 
different from the Vogel-Fulcher temperature obtained from the $\alpha$-relaxation time fit. 

To explore the fitting procedure in more depth, we also examined the dynamics of the
small and large particles separately. To this end we calculated $F_o^a(t)$ and
$\left< \delta r^{2(a)}(t) \right>$ where the calculations included sums over particles of type $a$, 
i.e.\ just the large or just the small particles. Then we defined a $\tau_\alpha^{a}$ as 
when $F_o^a(\tau_\alpha^a) = e^{-1}$ and obtained the diffusion coefficients from 
$\left< \delta r^{2(a)}(t) \right> = 6 D^a t$ at long times. Then we fitted mode-coupling power laws to 
$\tau_\alpha^a$ and $D^a$, four sets of data, and imposed that the fits have the same mode-coupling
temperature $T_c$. First, we imposed the restriction that $\tau_\alpha^a$ for both small and large 
particles have the same exponent $\gamma^{\tau}$ and that the exponent for the diffusion
coefficients $\gamma^D$  are the same, but $\gamma^{\tau}$ and $\gamma^D$ 
are allowed to be different. For this case, using the temperature range $12 \ge T \ge 6$ we obtained $\gamma^\tau = 2.9 \pm 0.1$
and $\gamma^D = 2.3 \pm 0.1$ with $T_c = 5.1 \pm 0.1$. 
Using the temperature range $12 \ge T \ge 7$ we obtained $\gamma^\tau = 2.5 \pm 0.1$
and $\gamma^D = 2.1 \pm 0.1$ with $T_c = 5.5 \pm 0.1$. 
Second, we allowed the exponents
to be different for the different types of particles as well as different for $D^a$ and $\tau_\alpha^a$. 
For these fits, using the temperature range $12 \ge T \ge 6$ we obtained $\gamma_s^\tau = 2.7 \pm 0.1$ and $\gamma_s^D = 1.9 \pm 0.1$ for
the small particles, $\gamma_l^\tau = 2.9 \pm 0.1$ and $\gamma_l^D = 2.4 \pm 0.1$ for the 
large particles, with a $T_c = 5.1 \pm 0.1$. Using the temperature range $12 \ge T \ge 7$ we obtained 
$\gamma_s^\tau = 2.3 \pm 0.1$ and $\gamma_s^D = 1.7 \pm 0.1$ for
the small particles, $\gamma_l^\tau = 2.4 \pm 0.1$ and $\gamma_l^D = 2.1 \pm 0.1$ for the 
large particles, with a $T_c = 5.7 \pm 0.1$.
The comparison of the $T_c$ values discussed in Sec. \ref{sec:dynamics} with the $T_c$ values presented in this paragraph  
shows that the value of the mode-coupling temperature depends only weakly 
on whether quantities pertaining to all particles or quantities pertaining to small and large particles are
used for fitting. 

In conclusion, for the system studied in this paper, we found that the mode-coupling temperature seems to
be somewhat ambiguous. For this reason, in this paper we used the notion of the mode-coupling temperature range
or $T_c$ range. On the basis of the fits discussed in Sec. \ref{sec:dynamics} and in this Appendix, we 
determined the $T_c$ range to be 5.6-5.1.

\section*{Appendix B}\label{sec:appendixB}
To aid in our fits to $S_4(q;\tau_\alpha)$ we calculate the ensemble independent dynamic
susceptibility 
\begin{equation}
\chi_4(t) = N^{-1} \left( \left< \sum_{n,m} w_n(t) w_m(t) \right> - \left<\sum_n w_n(t) \right>^2 \right), 
\end{equation}
which is the fluctuation in the overlap function $\sum_n w_n(t)$ where 
$w_n(t) = \Theta(a - \left| \mathbf{r}_n(t) - \mathbf{r}_n(0) \right| )$ and $\Theta(x)$
is Heavyside's step function.  Since $\chi_4(t)$
is a fluctuation quantity and some global fluctuations are suppressed in the $NVE$ and $NVT$ simulations
presented in this work, one cannot calculate $\chi_4(t)$ directly in an $NVE$ or $NVT$ simulation. 
By using methods developed previously \cite{Flenner:2011ht,Flenner2010prl,Lebowitz:1967vs} 
we account for these suppressed fluctuations, and
this gives us an added benefit of examining how dynamic heterogeneities may be related to experimentally
determinable quantities \cite{Berthier:2007jcp1,Berthier:2007jcp2,Berthier:2005te,DalleFerrier:2007kd,Brambilla:2009bz,CrausteThibierge:2010ki}. 

First we consider a transformation between the grand-canonical ensemble where the total number
of particles and the energy are allowed to fluctuate to the canonical ensemble where the energy
fluctuates, but the total number of particles are constant. In what follows we will 
denote $\left<\cdot\right>_{\mu V T}$ as an average in the grand-canonical ensemble, 
$\left<\cdot \right>_{NVT}$ as an average in the canonical ensemble, and $\left< \cdot \right>_{NVE}$
as an average in the micro-canonical ensemble. Note that there are two types of particles, thus there are two
chemical potentials $\mu_1$ and $\mu_2$, but we use a shorthand notation where $\mu$ represents both 
$\mu_1$ and $\mu_2$ and $N$ represents $N_1$ and $N_2$. Thus $\left< \cdot \right>_{NVT}$ 
is an average in an ensemble where $N_1$ and $N_2$ are not allowed to fluctuate.
Using the method of Lebowitz \textit{et al.}\ \cite{Lebowitz:1967vs}, we
obtain 
\begin{equation}\label{eq:chi1}
\left<N \right>_{\mu V T}  \chi_4(t) = \chi_4(t)|_{\delta N} + \chi_4(t)|_{NVT}
\end{equation}
where
\begin{eqnarray}\label{eq:chi2}
\chi_4(t)|_{\delta N}  &=& \left< (\delta N_1)^2 \right>_{\mu VT} 
\left[ \frac{\partial \left< \sum_n w_n \right>_{N V T}}{\partial N_1} \right]^2 \nonumber \\
&& + \left< (\delta N_2)^2 \right>_{\mu V T} 
\left[ \frac{\partial \left< \sum_n w_n \right>_{N V T}}{\partial N_2} \right]^2 \nonumber \\
&& + 2 \left< \delta N_1 \delta N_2 \right>_{\mu V T} \nonumber \\
&& \times \frac{\partial \left< \sum_n w_n \right>_{N V T}}{\partial N_1} 
\frac{\partial \left< \sum_n w_n \right>_{N V T}}{\partial N_2},
\end{eqnarray}
and $\chi_4(t)|_{NVT}$ is the fluctuation of $\sum_n w_n(t)$ calculated 
in the canonical ensemble. For the lowest two temperatures in this work, $T=5$, and
5.5 we used Eq.~\eqref{eq:chi1} to calculate $\chi_4(t)$. The temperatures $T \ge 6$ are discussed further below.

In practice, we convert the derivative
with respect to particle numbers to derivative with respect to concentration $c = N_1/N$
and volume fraction $\phi = 4 \pi (N_1 d_{1}^3 + N_2 d_{2}^3)/(3 V)$, where $d_1 = \sigma_{11}$ and $d_2 = \sigma_{22}$.  
This results in the following expression
\begin{eqnarray}
\chi_4(t)|_{\delta N} & =& \chi_\phi^2 H_1 + \chi_\phi \chi_c H_2 + \chi_c^2 H_3 \nonumber \\
&& + F_o(t)^2 H_4 + F_o(t) \chi_\phi H_5 + F_o(t) \chi_c H_6,
\nonumber \\
\end{eqnarray}
where $\chi_x = \partial F_o(t)/\partial x$. The $H_n$ are functions of 
$S_{nm} = \lim_{q\rightarrow0} S_{nm}(q)$ where 
\begin{equation}
S_{nm}(q) = (N_m N_n)^{-1/2} \left< \sum_{m,n} e^{i\mathbf{q} \cdot (\mathbf{r}_n - \mathbf{r}_m)}\right>
\end{equation} 
are the partial structure factors. The $H_n$ terms are given by
\begin{eqnarray}
H_1 & = & \left( \frac{\pi \rho}{6} \right)^2 \left[ d_1^6 x_1 S_{11} + 2 d_1^3 d_2^3 \sqrt{x_1 x_2} S_{12} + d_2^6 x_2 S_{22} 
\right] 
\nonumber \\ \label{H1}
\\
H_2  & = & \frac{\pi \rho}{3} \left[d_1^3 x_1 x_2 S_{11} - d_1^3 x_1 \sqrt{x_1 x_2} S_{12} \right.
\nonumber \\ \label{H2}
&& \left. + d_2^3 x_2 \sqrt{x_1 x_2} S_{12} - d_2^3 x_1 x_2 S_{22} \right] 
\\ \label{H3}
H_3 & = & x_2^2 x_1 S_{11} - 2 x_1 x_2 \sqrt{x_1 x_2} S_{12} + x_1^2 x_2 S_{22}
\\ \label{H4}
H_4 &= &x_1 S_{11} + 2 \sqrt{x_1 x_2} S_{12} + x_2 S_{22}
\\  
H_5 &=& \frac{\pi \rho}{3} \left[ d_1^3 x_1 S_{11} + (d_1^3 + d_2^3) \sqrt{x_1 x_2} S_{12} + d_2^3 x_2 S_{22} \right]
\nonumber \\ \label{H5} 
\\ 
H_6 &=& 2\left[ x_1 x_2 S_{11} + (x_2-x_1) \sqrt{x_1 x_2} S_{12} - x_1 x_2 S_{22} \right],
\nonumber \\
\label{H6}
\end{eqnarray}
where $\rho = N/V$ is the number density, $x_n = N_n/N$, and we have used that
\begin{eqnarray}
\left<N\right>_{\mu V T}^{-1} \left< \delta N_n \delta N_m \right>_{\mu V T} 
&=& \lim_{q \rightarrow 0} \sqrt{x_n x_m} S_{nm}(q) \nonumber \\ &=& \sqrt{x_n x_m} S_{nm}.
\end{eqnarray} 
In Eqs.~\eqref{H1} through \eqref{H6} the derivative with respect to volume fraction is performed at constant
concentration and the derivative with respect to concentration is performed at constant volume fraction. 
One can also perform the calculations varying the number density and the concentration
instead of the volume fraction and the concentration, but we found that this procedure involved  
large cancellations of terms of opposite sign. We also note that the $\chi_c^2$ term
could not be neglected in this work, but could be neglected in the analysis of earlier simulations of hard sphere systems \cite{Flenner:2011ht}. 
\begin{figure}
\includegraphics[width=3.2in]{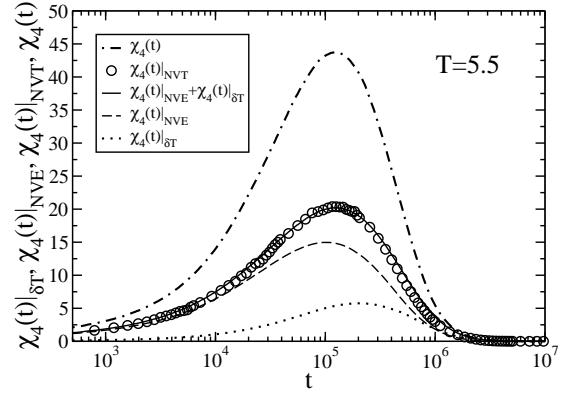}
\caption{\label{fig:correct}The four-point susceptibility calculated in an $NVT$ simulation 
$\chi_4(t)|_{NVT}$ (circles) compared to the susceptibility calculated in an $NVE$ simulation 
$\chi_4(t)|_{NVE}$ (dashed line), $\chi_4(t)_{\delta T}$ (dotted line), and the sum
$\chi_4(t)|_{NVE} + \chi_4(t)_{\delta T}$ (solid line). The dashed-dotted line is the full
ensemble independent susceptibility $\chi_4(t)$. 
}
\end{figure}
 
For $T \ge 6$ we performed only $NVE$ simulations, thus in Eq.~\ref{eq:chi1} we used 
$\chi_4(t)_{NVT} = \chi_4(t)_{NVE} + k_B T^2 \chi_T(t)^2/c_v$ where 
$c_v$ is the constant volume specific heat per particle and $\chi_4(t)|_{NVE}$ 
is the fluctuation of $\sum_n w_n(t)$ calculated in the $NVE$ simulation. To calculate all
the derivatives we performed at least two simulations at $x + \delta x$ and two at $x - \delta x$ 
where $x$ is $\phi$, $T$, or $c$.  The size of $\delta x$ depended on the temperature. 
 
We compared $\chi_4(t)_{NVT}$ and $\chi_4(t)_{NVE} + k_B T^2 \chi_T(t)^2/c_v$
and found excellent agreement for $T=5.5$, see Fig.~\ref{fig:correct}.  For comparison
we show the full ensemble independent susceptibility $\chi_4(t)$ as the dotted-dashed line
in Fig.~\ref{fig:correct}. 

We note that good agreement
between $\chi_4(t)_{NVT}$ and $\chi_4(t)_{NVE} + k_B T^2 \chi_T(t)^2/c_v$ does depend on 
the simulation parameters. Recall that we utilized a Nose-Hoover thermostat for the $NVT$ simulations 
and the LAMMPS simulation
package \cite{lammps} which utilizes the equations of motion of Shinoda \textit{et.~al} \cite{Shinoda:2004}. 
In the LAMMPS input script \cite{lammpsurl,lammpstd}, one has to provide a temperature damping parameter, {\it Tdamp}, which 
is related to the coupling to the heat bath. The {\it Tdamp} parameter is roughly equal to the time it takes
for the temperature to relax. A too small {\it Tdamp} and the temperature fluctuates 
wildly, but a too large {\it Tdamp} results in a large time for the temperature to equilibrate and 
a drift in the temperature for the very long simulations needed in this work. We found that a 
{\it Tdamp} of 2 time units to result in large fluctuations, thus a large  $\chi_4(t)|_{NVT}$. However,
noticeable temperature drift occurs after about several hundred million time steps for 
a {\it Tdamp} of 10,000 time units. We chose a {\it Tdamp} of 1,000 time units to minimize the effects of the
thermostat on the temperature fluctuations and to remove the energy drift. Note that we checked that
this choice of {\it Tdamp} did not influence the results at $T=5.5$ by comparing simulations of 
100$\tau_\alpha$ in the $NVE$ and $NVT$ ensembles. We also compared results of $NVE$ simulations of 10$\tau_\alpha$ 
and $NVT$ simulations of 100$\tau_\alpha$ for $T=5$ and found statistically the same results. However, the statistics are
poor for the shorter $NVE$ simulations and we needed the longer $NVT$ simulations to obtain sufficient statistics for this work. 

We fit $S_4(q;\tau_\alpha)$ using the value of $\chi_4(\tau_\alpha)$ calculated through the procedure described in this Appendix 
as $S_4(q=0;\tau_\alpha)$. This allows us to use slightly smaller systems and makes it possible to reach temperatures below the $T_c$ range. 
However, to examine what occurs without utilizing
$\chi_4(\tau_\alpha)$, we also fitted $S_4(q;\tau_\alpha)$ using only finite $q$ data. Thus, we found $\chi_4(\tau_\alpha)$ purely through
the extrapolation of $S_4(q;\tau_\alpha)$. These fits generally result in a lower value of $\chi_4(\tau_\alpha)$ and a smaller $\xi_4(\tau_\alpha)$,
especially when the correlation length $\xi_4(\tau_\alpha)$ and the susceptibility $\chi_4(\tau_\alpha)$
are large enough such that the plateau region of the Ornstein-Zernicke function is not visible in the finite $q$ data. 
However, the finite $q$ fits result in values that are are always within error of $\chi_4(t)$ determined from fits using also the $q=0$ point. 
Moreover, the correlation length $\xi_4(\tau_\alpha)$ obtained using only finite $a$ data
is close to, but systematically smaller than $\xi_4(\tau_\alpha)$ determined from fits using also the $q=0$ point. For most of the temperatures
the two values of $\xi_4(\tau_\alpha)$ are within error bars.

\begin{figure}
\includegraphics[width=3.2in]{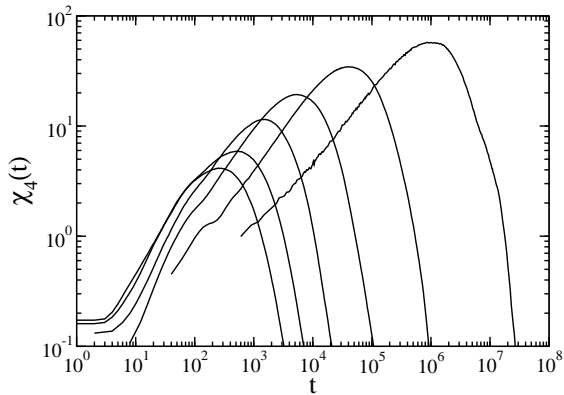}
\caption{\label{c4time}The ensemble independent susceptibility $\chi_4(t)$ for $T=10$, 9, 8, 7, 6, and 5 
listed from the smallest peak to the highest peak. These are the 
same temperatures as shown for $S_4(q;\tau_\alpha)$ in Fig.~\ref{fig:s4over}.}
\end{figure}

Shown in Fig.~\ref{c4time} is the full ensemble independent susceptibility $\chi_4(t)$ as a function of time for 
$T=10$, 9, 8, 7, 6, and 5. 
At $t=0$ $\chi_4(t)$ coincides with the small wave-vector limit of the total static structure factor,  
\begin{equation} 
\chi_4(t=0) = \lim_{q\to 0} N^{-1} \left< \sum_{m,n} e^{-i \mathbf{q} \cdot [\mathbf{r}_n(0) - \mathbf{r}_m(0)]} \right>  = \lim_{q\to 0} S(q).
\end{equation} 
$\chi_4(t)$ begins to grow during the $\beta$-relaxation and it peaks around $\tau_\alpha$. For longer times $\chi_4(t)$
decays to zero. The growth towards the peak can be well described by a power law 
$\chi_4(t) = \mathcal{X} + \chi_0 t^c$ where $c$ depends on temperature and decreases with decreasing temperature
in the temperature range studied. 


\end{document}